\documentstyle[12pt]{article}   
\begin{document}
\newcommand{\vev}[1]{\langle #1 \rangle}
\newcommand{\ltimes}{\times \hspace{-6pt} 
\raisebox{.2ex}{\mbox{$\scriptscriptstyle |$}} \;}
\def\mapright#1{\!\!\!\smash{
\mathop{\longrightarrow}\limits^{#1}}\!\!\!}
\newcommand{\bigoint}{\displaystyle \oint}
\newlength{\extraspace}  
\setlength{\extraspace}{2mm}
\newlength{\extraspaces}
\setlength{\extraspaces}{2.5mm}
\newcounter{dummy}
\newcommand{\be}{\begin{equation}
\addtolength{\abovedisplayskip}{\extraspaces}
\addtolength{\belowdisplayskip}{\extraspaces}
\addtolength{\abovedisplayshortskip}{\extraspace}
\addtolength{\belowdisplayshortskip}{\extraspace}}
\newcommand{\ee}{\end{equation}}
\newcommand{\ba}{\begin{eqnarray}
\addtolength{\abovedisplayskip}{\extraspaces}
\addtolength{\belowdisplayskip}{\extraspaces}
\addtolength{\abovedisplayshortskip}{\extraspace}
\addtolength{\belowdisplayshortskip}{\extraspace}}
\newcommand{\one}{{\bf 1}}
\newcommand{\ea}{\end{eqnarray}}
\newcommand{\R}{{\bf R}}
\newcommand{\Z}{{\bf Z}}
\newcommand{\tr}{{\rm tr}}
\newcommand{\N}{{\cal N}}
\newcommand{\Lbar}{\overline{L}}
\newcommand{\Nisf}{\N \! = \! 4}
\newcommand{\Nise}{\N \! = \! 8}
\newcommand{\cH}{{\cal H}}
\newcommand{\is}{\! &\! = \! & \! }
\newcommand{\nonu}{\nonumber\\[2.5mm]}
\renewcommand{\theequation}{\thesection.\arabic{equation}}
\renewcommand{\Im}{{\rm Im}}
\newcommand{\newsection}[1]{
\vspace{15mm}
\pagebreak[3]
\addtocounter{section}{1}
\setcounter{equation}{0}
\setcounter{subsection}{0}
\noindent
{\large\sc \thesection. #1}
\nopagebreak
\medskip
\nopagebreak}
\newcommand{\newsubsection}[1]{
\vspace{1cm}
\pagebreak[3]
\addtocounter{subsection}{1}
\addcontentsline{toc}{subsection}{\protect
\numberline{\arabic{section}.\arabic{subsection}}{#1}}
\noindent{ \it \thesubsection. #1}                 
\nopagebreak
\vspace{2mm}
\nopagebreak}
\noindent
\addtolength{\baselineskip}{.5mm}

\thispagestyle{empty}
\begin{flushright}
June 1997\\
{\sc hep-th}/9707179\\
{\sc itfa}-97/12
\end{flushright}
\vspace{1cm}
\thispagestyle{empty}
\begin{center}
{\large\sc Duality Symmetry of $\Nisf$ Yang-Mills Theory on $T^3$}
\\[17mm]
{\sc Feike Hacquebord}
and
{\sc Herman Verlinde}\\[5mm]
{\it Institute for Theoretical Physics}\\
{\it University of Amsterdam}\\
{\it Valckenierstraat 65, 1018 XE Amsterdam}
\\[24mm]
{\sc Abstract}
\end{center}
We study the spectrum of BPS states in $\Nisf$ supersymmetric $U(N)$
Yang-Mills theory. This theory has been proposed to describe $M$-theory on $T^3$
in the discrete light-cone formalism. We find that the degeneracy of irreducible
BPS bound states in this model exhibits a (partially hidden) $SL(5,\Z)$ duality
symmetry. Besides the electro-magnetic symmetry, this duality 
group also contains
Nahm-like transformations that interchange the rank $N$ of the gauge group with some
of the magnetic or electric fluxes. In the M-theory interpretation, this mapping
amounts to a reflection that interchanges the longitudinal direction with
one of the transverse directions.
\newpage

\newcommand{\ra}{\rightarrow}

\newsection{Introduction}

In this paper we will consider $\N\! =\! 4$ supersymmetric Yang-Mills 
theory in 3+1 dimensions. This model is described by 
the Lagrangian
\begin{eqnarray}
S & = & -\frac{1}{g^2}\int \! d^4x\, {\rm tr}\Bigl(\frac{1}{4}
F_{\mu\nu}^2+\frac{1}{2}(D_{\mu}X^I)^2 +
\psi\Gamma^i D_i \psi\nonumber\\
& & \qquad \qquad +\psi\Gamma^I[X^I,\psi]+\frac{1}{4}[X^I,X^J]^2 \Bigr) 
\end{eqnarray}
with $I=1,\ldots,6$, and can be thought of as the dimensional
reduction of $\N \! = \! 1$ SYM theory from ten to four dimensions.
Elegant semi-classical studies have revealed that the spectrum of 
dyonic BPS-saturated states in this theory exhibits an exact symmetry 
between electric and magnetic charges \cite{montonen-olive,Osborn,vw}.  
This $S$-duality symmetry 
is expected to extend into the full quantum regime, thereby providing 
an exact mapping between the strong coupling and weak coupling regions. 
Although the recent breakthroughs in non-perturbative supersymmetric
gauge 
theory and string theory have produced a substantial amount of evidence
for this duality conjecture, finding an explicit construction of this
duality
mapping still seems as difficult as ever.

The S-duality of the $\Nisf$ model forms an important ingredient
in the matrix theory formulation of 11-dimensional M-theory \cite{matrix}.
Matrix theory proposes a concrete identification 
between (the large $N$ limit of) the $U(N)$ SYM-model defined on a 
three-torus $T^3$ and type IIA or B superstring theory compactified 
on a two-torus $T^2$. 
This correspondence with perturbative string theory becomes most precise
in the limit when one of the directions of the three torus becomes much
larger than the others. In this IR limit one approaches an effective 
conformal field theory, that was argued in
\cite{motl,tomnati,matrixstring} to
correspond to a second quantized IIA string theory. A particularly
striking 
consequence of this conjectured correspondence is that the $S$-duality 
of the gauge model gets mapped to a simple $T$-duality in 
the string theory language \cite{susskind,wati}.

Central in the correspondence with M-theory on $T^3$ is the following 
construction of the 11-dimensional supersymmetry algebra in terms of the 
SYM  degrees of freedom \cite{mmatrix}.
The generators of the four-dimensional
$\Nisf$ supersymmetry algebra can be conveniently
combined into one single $SO(9,1)$ spinor supercharge $Q_\alpha$, by
considering the 4D SYM model as the 
dimensional reduction of ten-dimensional ${\cal N}=1$ SYM theory. 
In this ten dimensional notation   
the fermionic fields transform under the supersymmetry according to 
$\delta \psi= \Gamma^{\mu\nu}F_{\mu\nu}\epsilon$,
and the corresponding supercharge $Q_\alpha$ is equal to
\be
Q=\int_{T^3} \tr \left[\Gamma^r\Psi E_r
-\Gamma^0\Gamma^{rs}\Psi \frac{1}{2}F_{rs}\right].
\ee
Here and in the following the indices $r,s$ run from 1 to 9.
As the spatial part of our space-time manifold is compact we have 
an additional global supersymmetry: the action is invariant under adding
a constant spinor to $\psi$ via $\delta \psi = \tilde{\epsilon}$.
We denote the corresponding supercharge by $\tilde{Q}$
\be
\tilde{Q}= \int_{T^3} \! \tr\, \Psi,
\ee
where we put the volume of the three torus (and all its sides) equal to
one.
The supersymmetry algebra is
\begin{eqnarray}
\label{susy}
\left\{\tilde{Q}_\alpha , \tilde{Q}_\beta \right\} &=&
N\delta_{\alpha\beta}
\nonumber\\
\left\{Q_\alpha , \tilde{Q}_\beta\right\} &=& Z_{\alpha\beta}\\
\left\{\bar{Q}_\alpha , Q_\beta \right\} &=& 2 \Gamma^0 H +
2 \Gamma^i P_i \nonumber
\end{eqnarray} 
where the central charge term
\be
Z = \Gamma^0\Gamma^i e_i-\Gamma^{ij}m_{ij}
\ee
consists of the total electric and magnetic flux through the
three-torus,
defined via 
\ba
\label{flux}
e_i\is
\int_{T^3} \! \tr\, E_i,\nonu 
m_{ij}\is
\int_{T^3} \! \tr \, F_{ij}.
\ea
$H$ in (\ref{susy}) denotes the SYM Hamiltonian
and the quantities $P_i$ are the integrated energy momentum fluxes, defined 
(in 3+1 notation) via 
\be
P_i=\int_{T^3} \tr (E_jF_{ji}+\Pi^I D_i X_I+\frac{1}{2}i\Psi^{\rm
T}D_i\Psi),
\ee
where $\Pi_I$ defines the conjugate momentum to the Higgs scalar field $X_I$.
In writing the above supersymmetry algebra we have assumed that the $U(1)$ zero mode 
part of $\Pi_I$ vanishes. From the eleven dimensional M-theory perspective, this
means that we assume to be in the rest-frame in the uncompactified space directions.

In the next table we summarize the interpretation of the various fluxes
of $3\! + \! 1$-dimensional SYM theory in terms of charges of
11-dimensional
M-theory on $T^3$ and 10-dimensional IIA string theory compactified on
$T^2$. 
The latter correspondence, with the IIA string, proceeds by writing 
$T^3=T^2 \times S^1$ and interpreting the third direction (along the
$S^1$)
as the eleventh direction of M-theory. This leads to the following
list of correspondences (here $i,j$ run from 1 to 2)
\begin{eqnarray}
\begin{array}{ccccc}
  \begin{array}{|l|l|}
  \hline
  N & H \\[2mm]
  \hline
  e_3 & e_i\\[2mm]
  \hline
  m_3 & m_i\\[2mm]
  \hline
  p_3 & p_i\\[2mm]
  \hline
  \end{array}
 &\quad \leftrightarrow \quad &
  \begin{array}{|l|l|}
  \hline
  p_{+} & p_-\\[2mm]
  \hline
  p_3 & p_i\\[2mm]
  \hline
  m_{ij} & m_{3 j}\\[2mm]
  \hline
  m_{3+} & m_{i+}\\[2mm]
  \hline
  \end{array}
 &\quad \leftrightarrow \quad &
  \begin{array}{|l|l|}
  \hline
  p_{+} & p_- \\[2mm]
  \hline
  q_0 & p_i\\[2mm]
  \hline
  m_{ij} & w_j\\[2mm]
  \hline
  w_+ & m_{i+}\\[2mm]
  \hline
  \end{array}\\[1.9cm]
\mbox{4D SYM} && \mbox{M-theory on $T^3$} && \mbox{IIA String on $T^2$} 
\end{array}\nonumber
\end{eqnarray}

\noindent
Here $q_0$ denotes the D-particle number in the IIA string theory.
In the second and third table, the integers $m_{ij}$ -- corresponding to
the SYM magnetic fluxes -- denote the
wrapping numbers of the M-theory membrane around the $T^3$ and the
$D2$-brane
around the $T^2$ respectively. The M-theory membranes wrapped $m_{i3}$
times
around the 3-direction turn into IIA strings with winding number $w_i$.

States with non-zero momentum flux $P_i = p_i$ in the SYM theory
correspond
to membranes that are wrapped around the longitudinal +-direction.
In the original proposal of \cite{matrix}, M-theory arises in the limit
$N\ra \infty$ while restricting the spectrum of the $\Nisf$ SYM model to
the subspace of states that have energy of order $1/N$. This limit
amounts
to a decompactification of the longitudinal +-direction. States with
non-zero $p_i$ will thus correspond to infinite energy configurations 
containing membranes stretched along the light-cone direction. All
finite
energy states therefore must have $p_i=0$. 

An interesting proposal made in \cite{susskind-dlc} extends the
matrix theory conjecture to finite $N$, via a discrete light-cone
formalism. In this set-up, one compactifies the light-direction
on a circle and considers the truncation of the M-theory Hilbert
space to a given finite, discrete light-cone momentum identified
with the rank $N$. In this case, the longitudinal membranes are
no longer infinitely massive, and must be naturally included in
the spectrum.

The correspondence with string theory and M-theory gives a number
of predictions concerning the duality properties of the gauge theory.
These predictions in particular concern the BPS spectrum of the 
SYM model at finite $N$ as well as the behaviour of the model at large
$N$. For example, from the above table it is seen immediately that
electric-magnetic duality in the SYM theory indeed must follow from the 
T-duality on the two-torus, as first noted in \cite{susskind}. 
The T-duality that exchanges the D-particles with the D2-branes and
the KK momenta with the NS winding numbers in type IIA string theory, 
when translated to the first table indeed gives rise to the S-duality
that interchanges all the electric and magnetic fluxes. Via
this correspondence, the complete U-duality symmetry 
$SL(2,\Z) \times SL(3,\Z)$ of M-theory on $T^3$ is expected to be
realized as an exact symmetry of the matrix formalism \cite{susskind,wati}.

For finite $N$, however, there are reasons to suspect that the duality group
that acts on the BPS sector is in fact enlarged. Following the proposal
of \cite{susskind-dlc}, it should describe M-theory with more than 3 
directions compactified, and correspondingly one no longer needs to
restrict to states with vanishing momentum flux $p_i$. BPS states
at finite $N$ can therefore carry a total of 10 charges, labeled 
by $(N,e_i,m_i,p_i)$.  The goal of our study is to determine the
explicit form of the BPS degeneracies for finite $N$ as a function
of these charges, and thereby exhibit its full duality symmetry.
We will approach this problem in two ways: first from
M-theory and then directly from the $\Nisf$ SYM model. Our main finding is 
that the spectral degeneracy of individual bound states from both points of view
is indeed identical, and furthermore exhibits a full $SL(5,\Z)$
duality symmetry. Part of this large duality group also acts
on the rank $N$ of the gauge group. We will comment on the
geometric origin of this symmetry in the final section.

\newsection{BPS spectrum from M-theory}

The BPS-states that we will consider respect 1/4 of all supersymmetries.
Every such BPS-state in a fixed multiplet satisfies
\be
\left(\epsilon^\alpha Q_\alpha+\tilde{\epsilon}^\alpha 
\tilde{Q}_{\alpha}\right)
\left|{\rm BPS}\right>=0
\ee
for certain fixed collection of $SO(9,1)$ spinors $\epsilon$ and 
$\tilde{\epsilon}$. These spinors are, up to an overall factor, completely 
determined by the set of charges.

By taking the commutator with $\bar{Q}$ and $\tilde{Q}$ in the preceding
equation
we get two conditions for the spinors $\epsilon$ and $\tilde{\epsilon}$
\ba
2\Gamma^0H\epsilon
+2\Gamma^iP_i\epsilon
+\Gamma^0 Z \tilde{\epsilon}\is 0,\nonu
N \tilde{\epsilon}+Z^\dagger\epsilon \is 0.
\label{tepsilon}
\ea
Plugging $\tilde{\epsilon}$ in the first equation gives the following
equation for $\epsilon$
\be
\left(\Gamma^0H'+\Gamma^i P'_i\right)\epsilon=0
\label{conditie}
\ee
where $H'$ and $P_i'$ denote the Hamiltonian and momentum fluxes with
the zero-mode contributions removed. Explicitly,
\ba
H \is  \frac{1}{2N}(e_i^2 + m_{i}^2) + H' \nonu
P_i \is (e \wedge m)_i/N + P_i'.
\ea
{} From the last equation, we see that the eigenvalues of $P'_i$ are equal
to $p'_i = \kappa_i/N$ with 
\be
\label{kappa}
\kappa_i = Np_i - (e\wedge m)_i
\ee

The operator in equation (\ref{conditie})
should have eigenvalues equal to zero.
This is only the case when the magnitude of $H'$ and the length of $P'_i$
are equal, $H' = |P_i'|$. 
Using this relation, we can write equation (\ref{conditie}) 
in the following way
\be
\left(\Gamma_0+\Gamma^i\hat{\kappa}_i\right)\epsilon=0
\ee
where 
the vector $\hat{\kappa}_i$ denotes the unit vector in the direction of $\kappa_i$.

Now we are ready to determine the detailed BPS spectrum from discrete light-cone
M-theory. To this end it is useful to introduce the notion of an {\it irreducible}
BPS state, as a state that contains only one {\it single} BPS bound state.
Indeed, general BPS states can in principle combine more than one such bound state
into one second quantized BPS state. The combined state will still be BPS, provided
each of the irreducible constituent bound states all left invariant by the {\it same}
set of supersymmetries, {\it i.e.} with the same $\epsilon$ and $\tilde{\epsilon}$.

The degeneracy of irreducible BPS states is determined almost uniquely from U-duality
invariance, and from the known BPS spectrum of perturbative string states.
The relevant U-duality group for our case turns out to be as large as the
complete $SL(5,\Z)$ duality group of string theory on $T^3$, {\it i.e.} M-theory 
on $T^4$. At first sight, the appearance of this large symmetry may seem 
surprising, in particular since the supersymmetry algebra (\ref{susy}) itself
is not (manifestly) $SL(5,\Z)$  invariant. The origin of this large duality
group can be understood, however, by noting that the discrete light-cone version
of M-theory on $T^3$ can be obtained from M-theory on $T^4$ via a suitable limit,
essentially by applying an infinite boost to one of the compactified directions
of the $T^4$. The BPS spectrum should be invariant under such boosts.\footnote{
Alternatively, we can interpret the $U(N)$ SYM model as the low energy description 
of all possible bound states of $N$ D3 branes of IIB string theory on $T^3$. In 
this correspondence, $e_i$ denotes the NS string and $m_i$ the D-string winding 
number, while $p_i$ is the KK momentum. The SYM theory should thus encompass all 
BPS bound states of this system. It should be emphasized, however, that from
this reasoning we should only expect the BPS
{\it degeneracy} formula  to be $SL(5,\Z)$ symmetric, while of course the
energy spectrum is not, since the Yang-Mills model only represents a particular
limit of the $N$ D3 brane system.}

The 10 quantum numbers $(N,e_i,m_i,p_i)$ can be combined into a
$5\times 5$ anti-symmetric matrix on which $SL(5,\Z)$ acts by
simultaneous
left- and right-multiplication. The following bilinear combinations
\ba
\label{KI}
K_i = (Np_i - (e\wedge m)_i, \, p\! \cdot\! m\, , \, p\!\cdot\! e\, )
\ea
transform as 5 vector under this duality.
{} From this we deduce that the unique $SL(5,\Z)$ invariant scalar 
combination we can make out of the ten charges is the integral length of this
five-vector
\be
\label{KK}
|K| = \gcd(Np_i - (e\wedge m)_i, \, p\! \cdot\! m\, , \, p\! \cdot\! e\, )
\ee
Duality invariance thus predicts that the degeneracy of irreducible
BPS bound states should be expressible in terms of this quantity $|K|$ only.

To determine the explicit degeneracy formula, we note that by using
the U-duality symmetry, any irreducible bound state can be rotated into a
state that carries only KK momentum and NS string winding. Such a state must
necessarily be made up from a single fundamental IIA string. The invariant
$|K|$ for this perturbative string state simply reduces to the bilinear
combination of momenta and winding numbers that determines (via the BPS
restriction) the oscillator level of the string. The number of irreducible
BPS bound states is therefore counted by means of the chiral
string partition function 
\be
\sum_K c(K)q^K=(16)^2 \prod_n \left(\frac{1+q^n}{1-q^n}\right)^8
\label{generator}
\ee
with $K=|K|$ as defined in (\ref{KK}).

General BPS states of discrete light-cone gauge M-theory may consist of more
than one irreducible bound state. the total BPS condition requires that the
charges of these separate bound states must be compatible.
The complete second quantized partition sum is obtained by taking into
account all possible such ways of combining individual bound states into
a second quantized configuration with a given total charge. More detailed
comments on the combinatorial structure of the second quantized BPS
partition sum will be given in the next section.

\newsection{BPS-spectrum of $\Nisf$ SYM on  $T^3$}

The above description of the BPS spectrum can be reproduced directly
from the $U(N)$ gauge theory on $T^3$ as follows. First let us recall
the definition of the electro-magnetic flux quantum numbers.
To this end it is useful to decompose
the $U(N)$ gauge field into a trace and a traceless
part
\be
A_\mu 
=A^{U(1)}_\mu\one+A^{SU(N)}_\mu
\ee 
and allow the fields on $T^3$ to be periodic up to gauge
transformations of the form
\ba
A_i^{SU(N)}(x+a_j)\is \Omega_jA^{SU(N)}_i(x)\Omega_j^{-1}
\nonu \label{boundary}
A^{U(1)}_i(x+a_j)\is A^{U(1)}_i(x)-2\pi m_{ij}/N
\ea
with $m_{ij}$ integer. Here $a_i$ with $i=1,2,3$ denote the translation vectors
that define the three torus $T^3$.\footnote{We repeat here that for simplicity
we mostly take $T^3$ to be cubic with sides of length 1.}
The $SU(N)$ rotations $\Omega_i$ must satisfy the $\Z_N$ cocycle conditions
\be
\Omega_j\Omega_k = \Omega_k\Omega_j
e^{2\pi i \mu_{jk}/N}
\ee
for integer $\mu_{jk}$.
The quantities $\mu_{ij}/N$ define the 't Hooft $Z_N$ magnetic fluxes 
\cite{hooft} and $m_{ij}/N$ can be identified with the $U(1)$
magnetic flux defined in (\ref{flux}). 
The integers $\mu_{ij}$ and $m_{ij}$ are restricted 
via the condition that the combined gauge transformation (\ref{boundary})
defines a proper $U(N)$ rotation. This requirement translates into the 
Dirac quantization condition that the total flux $(\mu_{ij}-m_{ij})/N$
must be an integer. Hence $m_{ij} = \mu_{ij} (\mbox{mod $N$})$.

Electric flux carried by a given state is defined via the action of
quasi-periodic gauge rotations $\Omega[{\bf n}]$ defined via
\be
\Omega_j\Omega[{\bf n}] = \Omega[{\bf n}]\Omega_j
e^{2\pi i n_{j}/N}
\ee
with integer $n_j$. Such gauge rotations will preserve the boundary
condition (\ref{boundary}) on the gauge fields. A state $|\psi\rangle$ is defined 
to carry $SU(N)$ flux $\epsilon_j$ if it satisfies the eigenvalue condition
\be
\widehat{\Omega}[{\bf n}] | \psi \rangle = 
e^{{2\pi i \over N} n^j \epsilon_j} | \psi\rangle
\ee
where $\widehat{\Omega}[{\bf n}]$ denotes the quantum operator that implements the
gauge rotation $\Omega[{\bf n}]$ on the state $|\psi\rangle$. 
Similarly as for the magnetic flux, the electric
flux receives an overall $U(1)$ contribution $e_i$ defined in (\ref{flux}). 
The abelian
and non-abelian parts of the flux must again be related via $e_{i} = \epsilon_{i} 
(\mbox{mod $N$})$.

To determine the supersymmetric spectrum for a given set of charges, the idea is to
first reduce the phase space of the $U(N)$ SYM model to the space of classical
supersymmetric configurations, and then to quantize this BPS reduced phase space.
The justification for this procedure should come from the high degree of supersymmetry
in the problem, while furthermore the degeneracy of BPS states is known to be a very
robust quantity.

To obtain the reduced phase space, we recall that the SUSY transformation for the 
fermionic partners of the Yang-Mills fields reads (again using ten-dimensional
notation)
\be
\delta\Psi=\left(E_r\Gamma^{0r}+\frac{1}{2}F_{rs}\Gamma^{rs}\right)
\epsilon+\tilde{\epsilon}
\ee
The BPS restriction requires that the right-hand side vanishes for those $\epsilon$
and $\tilde{\epsilon}$ determined in the previous section. Thus in particular we can 
use the equation (\ref{tepsilon}) to express $\tilde{\epsilon}$ in terms of
$\epsilon$ and the $U(1)$ zero modes. The result is
\be
\delta\Psi=\left(E'_r\Gamma^{0r}+\frac{1}{2}F'_{rs}\Gamma^{rs}\right)
\epsilon=0.
\ee
where the primed quantities are equal to the un-primed ones with the constant
$U(1)$ parts removed. For BPS-states in a fixed multiplet, supersymmetry is 
unbroken for $\epsilon$ satisfying the equation (\ref{conditie}) above.
Hence for these BPS-states the following must hold for all spinors 
\be
\left(E'_r\Gamma^{0r}+\frac{1}{2}F'_{rs}\Gamma^{rs}\right)
\left(\Gamma^0-\Gamma^k\hat{\kappa}_k\right)\epsilon=0.
\ee
Note that $\Gamma^0-\Gamma^k\hat{\kappa}_k$ acts like a projection
operator on the space of spinors satisfying equation (\ref{conditie}).
We conclude that the matrix in spinor space in the last equation
has to vanish. This is the case when $E'$ and $F'$ satisfy the following
two conditions
\ba
\label{condition}
E'_i \hat{\kappa}^i\is 0\nonu
E'_{[r}\hat{\kappa}^{\phantom{'}}_{s]} \is F'_{rs}
\ea

{}From now one we shall omit the prime, and simply denote by $E$ and $F_{ij}$
the $U(N)$ fields without the $U(1)$ constant mode. We will also return to 
a 3+1-dimensional notation, and for additional notational convenience, use the $SL(3,\Z)$
symmetry to rotate the three-vector $\kappa_i$ defined in (\ref{kappa})
in the 3 direction. So we will choose coordinates such that
\ba
\label{kkk}
\qquad \qquad \qquad \kappa_3 \is Np_3 - e_i m_{i3} \\[2.5mm]
\qquad 
\qquad \qquad \kappa_j \is Np_j - e_i m_{ij} - e_3 m_{3j} \, = \, 0 \qquad 
\quad i,j = 1,2.
\nonumber
\ea
Here and from now on the indices $i,j$ run from 1 to 2.
{} From the second condition in (\ref{condition}) we read off that the gauge and Higgs
fields are flat on the plane perpendicular to $\hat{\kappa}$, meaning
\ba
\label{flat}
F_{ij} \is 0, \nonu
D_i X_J \is 0, \\[2mm]
[X_I,X_J]\is 0. \nonumber
\ea
In addition we have
\ba
\qquad E_i \is F_{3i}
\nonu
\Pi_I \is D_3X_I.
\ea
Finally, the first condition in (\ref{condition}) simply becomes
\be
E_3 = 0.
\ee
We can interpret this constraint equation as a gauge invariance condition under
arbitrary local shifts in the longitudinal gauge field $A_3$. We can therefore
exploit this gauge invariance by putting $A_3 = 0$. The relations 
(\ref{flat}) then simplify to the statement that the transversal
gauge fields $A_i$ and Higgs scalars $X_I$ satisfy the chiral 2D free field
equations
\ba
\label{chiral}
\partial_0 A_i=\partial_3A_i\nonu
\partial_0 X_I=\partial_3 X_I.
\ea
Thus we conclude that the BPS reduced theory is described by the left-moving
chiral sector of a two-dimensional sigma model with target space given by the
space of solutions to the flatness conditions (\ref{flat}), subject to the 
twisted boundary conditions specified by the electro-magnetic flux quantum numbers.

To determine the detailed properties of this sigma model, let us first consider
the case with all electro-magnetic fluxes equal to zero.
The only non-zero quantum numbers are therefore $N$ and $p_3$. 
In this case we can parameterize the space of solutions to (\ref{flat})
by means of the orbifold sigma model on the $N$-fold symmetric product space
\be
{(\R^6 \times T^2)^N\over S_N} \label{sym}
\ee
where $S_N$ denotes the permutation group of $N$ elements, acting on the
$N$ copies of the transversal space $T^2 \times \R^6$.\footnote{This orbifold
sigma model was first considered in relation with $\Nisf$ SYM theory on the 
three torus in \cite{reduc}.} To see that this is the right space, we observe that 
for solutions to (\ref{flat}) one can always choose a gauge in which
all $U(N)$ valued fields take the form of diagonal matrices. Each such
matrix field thus combines $N$ separate scalar fields, corresponding to
the $N$ eigenvalues. The $S_N$ permutation symmetry arises as a remnant of $U(N)$
gauge invariance, acting via its Weyl subgroup on the space of diagonal matrices.
Finally, the flat transversal gauge fields $A_i$ with $i=1,2$ give rise to
{\it periodic} 2D scalar fields, since constant shifts in $A_i$ by multiples
of $2\pi$ are pure gauge rotations.

The model thus reduces to the free limit of type IIA matrix string theory
\cite{motl,tomnati,matrixstring} in the discrete light-cone gauge \cite{susskind-dlc}.
As explained in \cite{matrixstring}, the Hilbert space of the model decomposes into
twisted sectors labeled by the partitions of $N$, in which the eigenvalue fields combine
into a collection of `long strings' of individual length $n_k$ such that the total length
adds up to $\sum_k n_k = N$.
Each such string is made up from, say, $n_k$ eigenvalues that, by their periodicity
condition around the 3-direction are connected via a cyclic permutation of order $n_k$.
In the M-theory interpretation, all these separate strings will indeed correspond to 
separate bound states, {\it i.e.} particles that each can move independently in the
uncompactified space directions. The general form of the BPS partition function of
symmetric product sigma models of the form (\ref{sym}) has been described in detail
in \cite{dmvv}.


In the following we will mainly concentrate on the irreducible states, describing
one single BPS particle. These necessarily consist of one single string of maximal
length. In the present case, with zero total electro-magnetic flux, this
maximal string has total winding number $N$ around the 3-direction.
Correspondingly, its oscillation modes have energies that are quantized in
units of $1/N$. Thus the degeneracy of states as a function of $N$ and $p_3$
is obtained by evaluating the chiral superstring partition function,
as given in (\ref{generator}), at oscillator level $Np_3$.

Next let us turn on the magnetic flux $m_3$. The space of solutions to (\ref{flat})
with twisted boundary conditions (\ref{boundary}) around the transverse torus
again takes the form of a symmetric product
\be
\label{symp}
{(\R^6 \times T^2)^{N'}\over S_{N'}}
\ee
but where now $N' = \gcd(N,m_3)$. In order to visualize this reduction,
we note that gauge rotations $\Omega_i$ with $\Omega_1 \Omega_2 = \Omega_2\Omega_1
\exp(2\pi i m_3/N)$, that define the twisted boundary conditions, can be chosen to
lie within an $SU(k)$ subfactor of $U(N)$ where $k= N/\gcd(N,m_3)$. By
decomposing the matrix valued fields according to the action of $SU(k) \otimes U(N')$
with $N' = \gcd(N,m_3)$, we can thus factor out a sector of $U(N')$ valued
field variables that are unaffected by the twisted boundary conditions.
Now following the same reasoning as before, these fields parameterize the symmetric
product space of the above form. Note that in the particular case that $m_3 =1$,
the whole $SU(N)$ part of the moduli space of solution to (\ref{flat}) collapses
to a point, so that only the $U(1)$ part survives.

By a very similar reasoning we can also include the electric flux $e_3$ in our 
description. Like with the magnetic flux, an electric flux $e_3=1$ has the effect 
of reducing the $SU(N)$ part of the vacuum moduli space (\ref{flat}) to a point,
or rather, it projects out just one single supersymmetric state in the $SU(N)$
sector \cite{witten}. More generally, however, it can be seen that one can
again factor out a $U(N')$ subfactor of the model, that is unaffected by
both the electric and magnetic flux $e_3$ and $m_3$, where now
\be
\label{Np}
N' = \gcd(N,m_3,e_3).
\ee
The BPS sector for non-zero $m_3$ and $e_3$ is thus obtained by quantizing the
supersymmetric orbifold sigma model on (\ref{symp}), with $N'$ equal to (\ref{Np}).

The spectrum of irreducible BPS bound states is obtained as before, by considering 
the Hilbert space sector defined by the eigenvalue string of maximal length. The
maximal winding number is now equal to $N'$. The total momentum along this string is
determined by the remaining quantum numbers of the BPS state, and should be equal
to
\be
\label{pp}
p_3 - e_im_{i3}/N,
\ee
which is the total momentum of the BPS state minus the contribution from the
$U(1)$ electro-magnetic fluxes. Notice that the latter contribution is in general
fractional. However, since the oscillation modes of the long string states also
have fractional oscillation number quantized in units of $1/N'$, this fractional 
total momentum (\ref{pp}) can in fact be obtained via integer string oscillation
levels. The total oscillation level corresponding to this momentum flux is
\be
\label{kn}
K = N' \times (p_3 - e_im_{i3}/N) 
\ee
and it is easy to check using (\ref{kkk}) that this is an integer. In fact, after 
taking the direction of $\kappa_i$ again arbitrary, we find that the integer quantity
(\ref{kn}) becomes equal to the $SL(5,\Z)$ invariant length
$|K|$ defined in (\ref{KK}) of the five-vector (\ref{KI}). In this way we 
reproduce the description of the BPS spectrum given in the previous section.
In particular we find confirmation that the degeneracy of supersymmetric 
bound states has a discrete $SL(5,\Z)$ duality invariance.

As mentioned already, the complete supersymmetric spectrum of the $\Nisf$
SYM model contains many more sectors, that in M-theory describe configurations 
of multiple BPS bound states. These correspond to the other twisted sectors
in the orbifold model (\ref{symp}), describing multiple strings with separate
lengths $n_k$ with $\sum_k n_k = N'$. Via their zero modes, these strings
each separately carry all the possible flux quantum numbers. The degeneracy 
of these separate states as a function of these charges is identical to the 
$SL(5,\Z)$ invariant result just described. However, it turns out that the 
combinatorics by which many such states can be combined into one second 
quantized BPS state no longer respects the full $SL(5,\Z)$ symmetry. For
this it would be necessary that, via the compatibility of the individual 
BPS conditions, the 10 dimensional charge vectors of all constituent states 
must align in the same direction. It can be seen, however, that this alignment 
is not entirely implied: in the above notation, all charges must indeed align,
{\it except} for the individual $p_3$ momenta. It appears therefore that the
full second quantized BPS partition sum exhibits a somewhat smaller duality 
symmetry than $SL(5,\Z)$. The symmetry that gets preserved, however, is still 
substantially bigger than the $SL(2,\Z)\times SL(3,Z)$ U-duality of M-theory 
on $T^3$. In the next section we will describe some of these extra symmetries
in more detail.

\newsection{Nahm Duality}

An especially interesting class of duality transformations that act on 
BPS sectors of $U(N)$ SYM theory are those that interchange the rank $N$
of the gauge group with the magnetic flux. These type of
transformations are very similar to the Nahm-type transformations,
considered e.g. in \cite{braam}. In the notation used above (with
$\hat\kappa$ rotated in the 3-direction) the BPS reduced quantum
phase space exhibits a manifest symmetry under the interchange
\be
N \leftrightarrow m_{3} \hspace{1cm}
m_i \rightarrow \epsilon^{ij} m_j \hspace{1cm}
e_i \leftrightarrow p_i, \label{symmetry1}
\ee
as well as under its electromagnetic dual counterpart
\be
N \leftrightarrow e_3 \hspace{1cm}
e_i \rightarrow \epsilon^{ij} e_j \hspace{1cm}
m_i \leftrightarrow p_i.  \label{symmetry2}
\ee
In particular one can verify that the relations (\ref{kkk}) as well as the
integer $N'$ are invariant under these two mappings. 
This last type of duality symmetry is particularly interesting, because from
M-theory, the symmetry (\ref{symmetry2}) must be some manifestation of
11-dimensional covariance. Because of this possible relevance, it is of interest 
to know whether the symmetry perhaps extends to the full $\Nisf$ model.
Although we do not directly expect that this is the case, we cannot resist
here to review the geometrical origin of this duality mapping, which
in fact can be defined for arbitrary non-BPS gauge configurations.

\newcommand{\Aa}{A}

\newcommand{\DAa}{{D}}
\newcommand{\hhat}{\widehat}

Consider an arbitrary $U(N)$ gauge field 
$A=A_x+iA_y$ on $T^2$ with magnetic flux $M$.
This configuration can be related to a dual $U(M)$ gauge field  with flux $N$ 
defined on the dual torus $\hat{T}^2$ as follows. The key idea of the construction 
is to consider the parameter family of connections on $T^2$ of the form 
(here $a,b$ denote the $U(N)$ color indices)
\be
\Aa^{ab}(x;z) =  A^{ab} (x)  +  2\pi {\bf 1}^{ab} z
\ee
with $z=z_1+iz_2$ a complex coordinate 
on the dual torus $\hhat{T}^2$. 
The mapping proceeds by considering the space of zero modes of the Dirac-Weyl
equation defined by $A$. Let $\DAa$ and $\DAa^*$ denote the corresponding Dirac operator 
acting on right- and left-moving spinors, respectively.
We will assume that $\DAa^*$ has no zero-modes. According to the index theorem, 
the number of zero-modes of $D$ then equals the magnetic flux $M$ 
\be
\label{one}
\DAa\psi_i(x;z) = 0
\ee
with $i=1,\ldots,M.$ We can choose an orthonormal basis of zero modes, satisfying 
\be
\int \! d^2 x \; \overline{\psi}_i(x;z)\psi_j(x;z) = \delta_{ij}
\ee
The dual $U(M)$ gauge field $\hhat{\Aa}$ on $\hhat{T}^2$ is now defined as 
\be
\hhat{A}_{ij}(z)  =  
\int \! d^2 x \; 
\overline{\psi}^a_i(x;z) 
\hhat\partial \psi_{aj}(x;z)
\ee
with $\hhat\partial  = {\partial\over\partial z }$.
This definition is equivalent to the following formula
for the dual covariant derivative $\hhat \DAa$
\be
\label{defn}
\hhat{\DAa}_\alpha \psi = 
({\bf 1}- {\bf P}) \hhat\partial_\alpha  \psi 
\ee
where ${\bf P}$ denotes the projection on the space of zero 
modes $\psi_i$ of the operator $D$. 
It can be shown that this
dual gauge field has magnetic flux equal to $N$, and moreover
that the mapping from $A$ to $\hhat{A}$ is a true duality,
in that it squares to the identity.

To make these properties more manifest, it is useful to obtain a somewhat 
more explicit form of the dual covariant derivative $\hhat{D}_\mu$.
To this end, define the Green function
\be
\Delta G(x,y) = \delta^{(2)}(x-y)
\ee
of the Laplacian $\Delta = D D^* + D^* D$,  and introduce the notation
\be
({\bf G}\psi)(x;z) = \int \! d^2y \; G(x,y) \psi(y;z).
\ee
Then the projection operator ${\bf P}$ satisfies the relation 
\be
\label{pgreen}
{\bf 1} - {\bf P}=  D^* {\bf G} D. 
\ee
Inserting this identity into the definition (\ref{defn}), we find that 
\be
\hhat{\DAa}^* {\psi} = D^* {\bf G} D \hat{\partial}^* {\psi}  = 0 
\label{two}
\ee
since $[D,\hat{\partial}^*] = 0$.
Hence the zero modes of the Dirac-Weyl operator on the dual torus are
equal to the ones on the original torus, only with the opposite
chirality. By an explicit construction \cite{christiaan} one can indeed 
verify that the dual $U(M)$ gauge field has magnetic flux equal to $N$,
as predicted by the index theorem. 
Thus the Nahm transformation can be summarized in an elegant way 
by means of the two equations (\ref{one}) and (\ref{two}),
which combined indeed completely specify the map from $A$ to $\hhat{A}$.
Note in particular that in this form, one of the `magical' properties of the Nahm
transformation has become manifest, namely that it is a map of order
2, {\it i.e.} it squares to the identity:
\be
\hhat{\hhat A} = A,
\ee  
up to gauge equivalence.  Notice further that the mapping
is defined for arbitrary connections $A$.

When we translate the above construction back to our 3+1-dimensional setting,
the resulting mapping indeed interchanges $N$ and $m_3$ as advocated. In addition,
since it is a mapping from $T^2$ to the dual torus $\hhat{T}^2$, the definition
of the momentum flux in the direction of the $T^2$ gets interchanged with that
of the electric flux along this direction, as indicated in (\ref{symmetry2}). 
Naturally, the electric flux represents the conjugate momentum to the constant
mode of the gauge fields $A_i$, and thus indeed defines the total momentum on
the dual torus $\hhat{T}^2$. Further inspection also shows that the magnetic
fluxes $m_i$ get reflected, as predicted.

Finally, we should of course note that in the interpretation of the $U(N)$
SYM model as describing $N$ D3 branes, this Nahm duality is nothing other 
than a double T-duality along the 1-2 directions of the three torus.
\footnote{To recognize this interpretation of the mapping (\ref{symmetry1}), 
see the translation code summarized in one of the previous footnotes. 
This interpretation of the Nahm transformation, as related 
to T-duality in string theory, was first suggested in \cite{mikegreg}.} 
{} From this perspective, it in fact seems somewhat 
surprising that the Yang-Mills theory (via the Nahm transformation) 
still knows about this T-duality symmetry, in spite of the fact that it arises
from string theory via the zero-slope limit.

\medskip
\medskip

\medskip

{\noindent \sc Acknowledgements}

\noindent
The initial stages of this work were done in collaboration with Robbert
Dijkgraaf, Christiaan Hofman, and Erik Verlinde. We further acknowledge
valuable discussions with R. Dijkgraaf, C. Hofman, J-S. Park, S. Ramgoolam, 
W. Taylor, and E. Verlinde. This research is partly supported by a Pionier
Fellowship of NWO, a Fellowship of the Royal Dutch Academy of Sciences
(K.N.A.W.), the Packard Foundation and the A.P. Sloan Foundation.

\vspace{15mm}
\pagebreak[3]

\renewcommand{\Large}{\large}


\begin{thebibliography}{99}
\addtolength{\itemsep}{-6pt}


\bibitem{montonen-olive}
C. Montonen and D. Olive, ``Magnetic monopoles as gauge particles,'' 
Phys. Lett. {\bf 72B} (1977) 117; P. Goddard, J. Nuyts and D. Olive, 
``Gauge theories and magnetic charge,'' Nucl. Phys. {\bf B125} (1977) 1.
\bibitem{Osborn} 
H. Osborn, ``Topological charges for $N=4$ supersymmetric
gauge theories and monopoles of spin 1,'' Phys. Lett. {\bf 83B} (1979)
321;
A. Sen, Phys. Lett. {\bf B329} (1994) 217, {\tt hep-th/9402032}. 


\bibitem{vw}
C. Vafa and E. Witten, ``A strong coupling test
of S-duality, '' Nucl. Phys. {\bf B431} (1994) 3, {\tt hep-th/9408074};
L. Girardello, A. Giveon, M. Porrati, A. Zaffaroni, Nucl. Phys. 
{\bf B448} (1995) 127, {\tt hep-th/9502057}.

\bibitem{matrix} T. Banks, W. Fischler, S.H. Shenker and L. Susskind, 
``M theory as a Matrix Model: a Conjecture,'' Phys. Rev. {\bf D55}
(1997) 5112, 
{\tt hep-th/9610043}.

\bibitem{susskind} L. Susskind, ``T-duality in M(atrix) Theory and 
S-duality in Field Theory,'' {\tt hep-th/9611164}.

\bibitem{wati}
W. Taylor, ``D-brane Field Theory on Compact Spaces,''
Phys. Lett. {\bf B394} (1997) 283, {\tt hep-th/9611042};
O. Ganor, S. Ramgoolam and W. Taylor, ``Branes, fluxes and duality in 
M(atrix) theory,'' Nucl. Phys. {\bf B492} (1997) 191, {\tt
hep-th/9611202}.

\bibitem{motl} 
L. Motl, ``Proposals on Nonperturbative Superstring Interactions,''
{\tt hep-th/9701025}.

\bibitem{tomnati} 
T. Banks and N. Seiberg, ``Strings from Matrices,'' {\tt
hep-th/9702187}. 

\bibitem{matrixstring} 
R. Dijkgraaf, E. Verlinde, and H. Verlinde, ``Matrix String Theory,'' 
{\tt hep-th/9703030}.

\bibitem{mmatrix} T. Banks, N. Seiberg, and S. Shenker, ``Branes from Matrices,''
Nucl. Phys {\bf B490} (1997) 91, {\tt hep-th/9612157}.

\bibitem{susskind-dlc} 
L. Susskind, ``Another Conjecture about M(atrix) theory,''
{\tt hep-th/9704080}.

\bibitem{hooft} 
G. 't Hooft, ``A property of Electric and Magnetic Flux in Non-Abelian Gauge Theories,'' Nucl. Phys. {\bf B153} (1979) 141.

\bibitem{reduc}
J. Harvey, G. Moore and A. Strominger, ``Reducing $S$-duality
to $T$-duality,'' Phys. Rev. {\bf D52} (1995) 7161, 
{\tt hep-th/9501022}. 

\bibitem{dmvv}
R. Dijkgraaf, G. Moore, E. Verlinde and H. Verlinde, ``Elliptic genera
of Symmetric Products and Second Quantized Srings,'' 
Commun. Math. Phys. {\bf 185} (1997) 197, {\tt hep-th/9608096}.

\bibitem{witten} 
E. Witten, ``Bound states of Strings and $p$-Branes,''
Nucl. Phys. {\bf B460} (1996) 335, {\tt hep-th/9510135}.

\bibitem{braam}
P. J. Braam and P. van Baal, ``Nahm's transformation for Instantons,''
 Commun. Math. Phys. {\bf 122} (1989) 267; 
P. van Baal, ``Instanton moduli for $T^3\times \R$,''
Nucl. Phys. Proc. Suppl. 49 (1996) 238, {\tt hep-th/9512223}.

\bibitem{christiaan} 
C. Hofman and E. Verlinde, unpublished.

\bibitem{mikegreg} M. R. Douglas and G. Moore, ``D-branes, Quivers, and 
A L E Instantons,''
{\tt hep-th/9603167}.
\end{thebibliography}
\end{document}